\documentclass{sf2a-conf2017}
\usepackage{graphicx}
\usepackage{hyperref}
\usepackage[]{natbib}  
\usepackage{epstopdf}
\usepackage{color}
\usepackage[utf8]{inputenc}

\def\BibTeX{{\rm B\kern-.05em{\sc i\kern-.025em b}\kern-.08em
    T\kern-.1667em\lower.7ex\hbox{E}\kern-.125emX}}
\bibpunct{(}{)}{;}{a}{}{,}  %%%%%%%%%%%%%  A&A bibliography style

\newcommand{\ind}[1]{_{\mathrm{#1}}}

\newcommand\DPi{\Delta\Pi\ind{1}}
\newcommand\Dnu{\Delta\nu}

\newcommand\dnurotcore{\delta\nu\ind{rot, core}}

\newcommand\Dtaum{\Delta\tau\ind{m}}

\newcommand\numax{\nu\ind{max}}
\newcommand\N{\mathcal{N}}

\newcommand{\Omegacore}{\Omega\ind{core}}
\newcommand{\Msol}{M\ind{\odot}}
\newcommand\Teff{T\ind{eff}}

\newcommand\numaxsol{{\nu\ind{max,\odot}}}
\newcommand\Dnusol{\Dnu\ind{\odot}}
\newcommand\Tsol{{T\ind{\odot}}}

\begin{document}

\TitreGlobal{SF2A 2017}

%%-----------------------------------------------------------------
%%      the top matter
%%

\title{Large-scale measurements of the red giant core rotation through asteroseismology}

\runningtitle{Large-scale measurements of the red giant core rotation through asteroseismology}

\author{C. Gehan}\address{Observatoire de Paris, LESIA, Université Pierre et Marie Curie, Université Paris Diderot, PSL}

%% IF Author3 has the same affiliation than Author1:
\author{B. Mosser$^1$}
\author{E. Michel$^1$}

%% Keep this line, even if the page will be settled afterwards.
\setcounter{page}{237}

%%-----------------------------------------------------------------

\maketitle

%%-----------------------------------------------------------------
%%        The abstract
%% 
%%  Warning!  within the abstract:
%%  - do not use macros. 
%%  - do not use commands like: \cite, \citet, \citep ... etc.

\begin{abstract}
Red giant stars are solar-like pulsators presenting mixed-modes. Such modes consist in a coupling between pressure waves propagating in the external convective envelope and gravity waves propagating in the radiative interior. Therefore, the red giant asteroseismology provides us with a direct view on their core and opens the possibility to monitor the evolution of their core rotation. Previous measurements of the mean core rotation revealed that angular momentum is efficiently transferred from the core to the envelope inside red giants, but the physical mechanisms at work are not yet fully understood. We thus need stronger observational constraints on the evolution of the red giant core rotation.
\newline
In this context, we developed an automated method to determine the mean core rotation of red giant branch stars observed with \textit{Kepler}. This automated method is paving the way for the future PLATO data, representing hundreds of thousands of potential red giant oscillation spectra.
\newline
Results obtained for almost 1200 red giant branch stars indicate that the rate of the core rotation braking is lower than previoulsy estimated and does not seem to depend on the stellar mass.
\end{abstract}

%% Insert the keywords (to appear in the ADS indexing)
%% Keywords must be separated by a comma
\begin{keywords}
Stars: oscillations, - Stars: interior, - Stars: evolution, - Stars: rotation
\end{keywords}

%%-----------------------------------------------------------------

\section{Introduction}

%%---------------------

The \textit{Kepler} NASA space mission has provided frequency oscillation spectra of unprecedent quality, opening the possibility to study the red giant core rotation.
Red giants are evolved intermediate-mass stars, between 0.5 and 8 $\Msol$. Hydrogen is exhausted in the core and is burning in a shell above the core. Red giants can lay in two evolutionary phases, the red giant branch (RGB) or the red clump. During the RGB phase, the star undergoes strong structural changes: the inert helium core contracts while the convective envelope deepens and extends, leading to a strong increase of the stellar radius and luminosity. The clump phase comes when the red giant starts burning helium in the core.
\newline
Red giant stars are solar-type pulsators: internal seismic waves are stochastically excited by the turbulent convection in the external envelope \citep{De_Ridder}. They present mixed-modes, consisting in a coupling between pressure waves propagating in the convective envelope and gravity waves propagating in the radiative interior \citep{Beck}. This feature allows us to have a direct insight on their core, which is not the case for solar-type pulsators on the main sequence like the Sun.
\newline
Rotation strongly impacts the stellar structure and evolution by altering the chemical element mixing inside stars \citep{Lagarde}, but including rotation in stellar evolution models is still challenging. Models still predict central rotation rates at least ten times too large compared to asteroseismic measurements. It is thus of prime importance to know how internal rotation evolves in time. Moreover, measurements of the mean core rotation of almost 300 red giants have revealed that the red giant core is slowing down along the RGB, while the core is contracting \citep{Mosser_2012c}. This necessarily implies that a very efficient angular momentum transport from the core to the envelope is at work inside red giants. Magnetic fields, \citep{Cantiello}, mixed-modes \citep{Belkacem_2015a, Belkacem_2015b} or internal gravity waves \citep{Fuller, Pincon} stand among the physical mechanisms that can dissipate angular momentum. Nevertheless, no modelling based on such physical mechanisms can reproduce the measured orders of magnitude of the red giant core rotation. We thus need stronger observational constraints on the evolution of rotation inside evolved stars.

%%-------------------------

\section{Automatic measurements of the red giant mean core rotation}

%%-------------------------

\begin{figure*}
\centering
\includegraphics[width=8.cm]{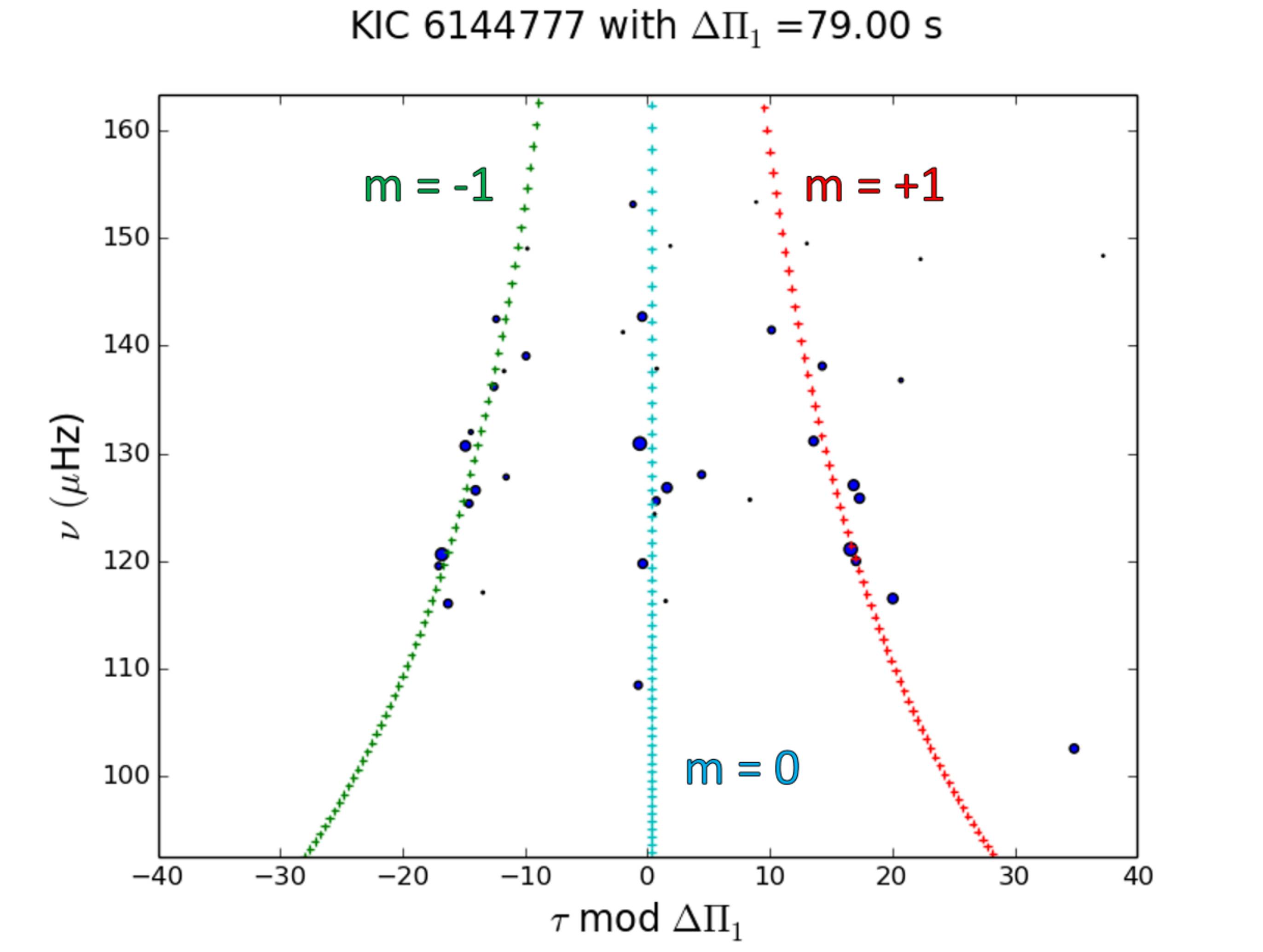}
\caption{Stretched period échelle diagram for a RGB star with an intermediate inclination angle. The rotational components are identified in an automatic way with a correlation of the observed spectrum with a synthetic one through Eq.~\ref{eqt-tau-dnurot}. The color codes the azimuthal order: the $m=\lbrace-1,0,1\rbrace$ rotational components are represented in green, light blue and red respectively.}
\label{fig-echelle}
\end{figure*}

Rotation has an effect on oscillation spectra which is similar to the Zeeman effect of a magnetic field on the energy levels of an atom: it lifts the degeneracy between oscillation modes having the same angular degree $\ell$ but different azimuthal orders $m$. 
We consider in this study dipole mixed-modes with a degree $\ell=1$ as they mostly probe the red giant core \citep{Goupil}. In these conditions rotation generates three rotational components having azimuthal order values $m \in \lbrace -1,0,1 \rbrace$.
\newline
We can build échelle diagrams based on stretched periods $\tau$ that reveal the rotational components \citep{Mosser_2015, Gehan_2016a, Gehan_2016b}. The number of visible rotational components depends on the stellar inclination: the two components associated to $m=\pm1$ are present when the star is seen equator-on, all the three rotational components are visible for intermediate inclination configurations, and only the rotational component associated to $m=0$ can be observed when the star is seen pole-on.
\newline
The stretched period spacing between dipole mixed-modes having the same azimuthal order depends on the mean core rotational splitting $\dnurotcore$ \citep{Mosser_2015}. It varies as
\begin{equation}\label{eqt-tau-dnurot}
 \Dtaum = \DPi \left( 1 + 2 \, m \, \zeta \, \frac{\dnurotcore}{\nu} \right),
\end{equation}
where $\DPi$ is the asymptotic period spacing of pure dipole gravity modes, $\zeta$ defines the stretching of the mode periods, and $\nu$ are the mode frequencies. The measurement of $\dnurotcore$ gives an estimate of the mean core angular velocity \citep{Goupil, Mosser_2015} through
\begin{equation}
\dnurotcore \simeq \frac{1}{2} \left\langle \frac{\Omegacore}{2 \pi} \right\rangle.
\end{equation}
We developed a method allowing an automated identification of the rotational components based on the correlation of the observed spectrum with synthetic ones constructed through Eq.\ref{eqt-tau-dnurot} \citep{Gehan_2017}.

%----------------------------

\section{Constraining the rate of the core braking} 

%%-------------------------

\begin{figure*}
\centering
\includegraphics[width=12cm]{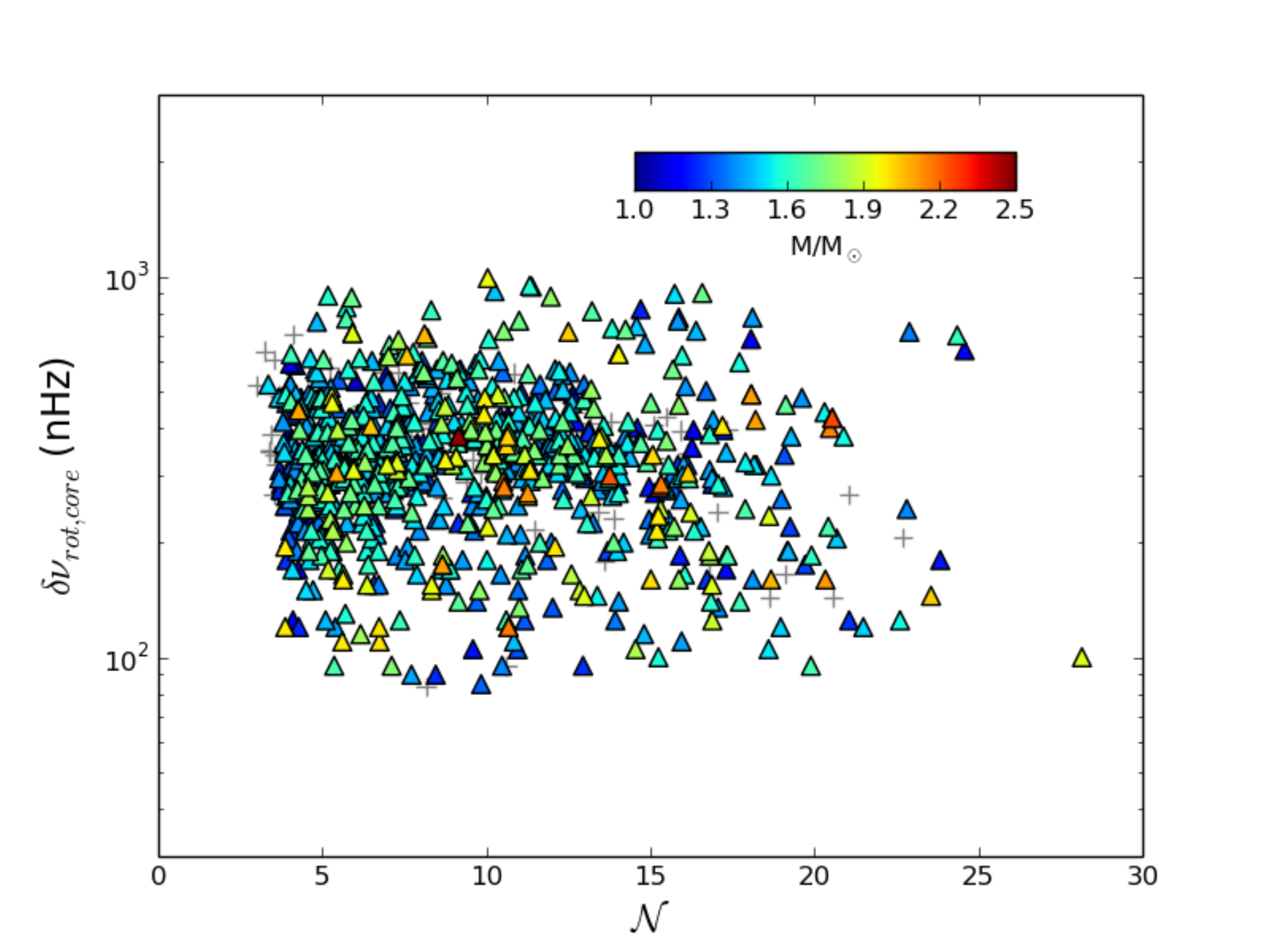}
\caption{Mean core rotational splitting as a function of the mixed-mode density, in semi-log scale. Gray crosses represent the measurements of \cite{Mosser_2012c}. Colored triangles represent the measurements obtained in this study \citep{Gehan_2017}, the color coding the mass estimated from the asteroseismic global parameters. The mixed-mode density increases along the evolution on the red giant branch.}
\label{fig-dnurot}
\end{figure*}

\begin{figure*}[h]
\centering
\includegraphics[width=12cm]{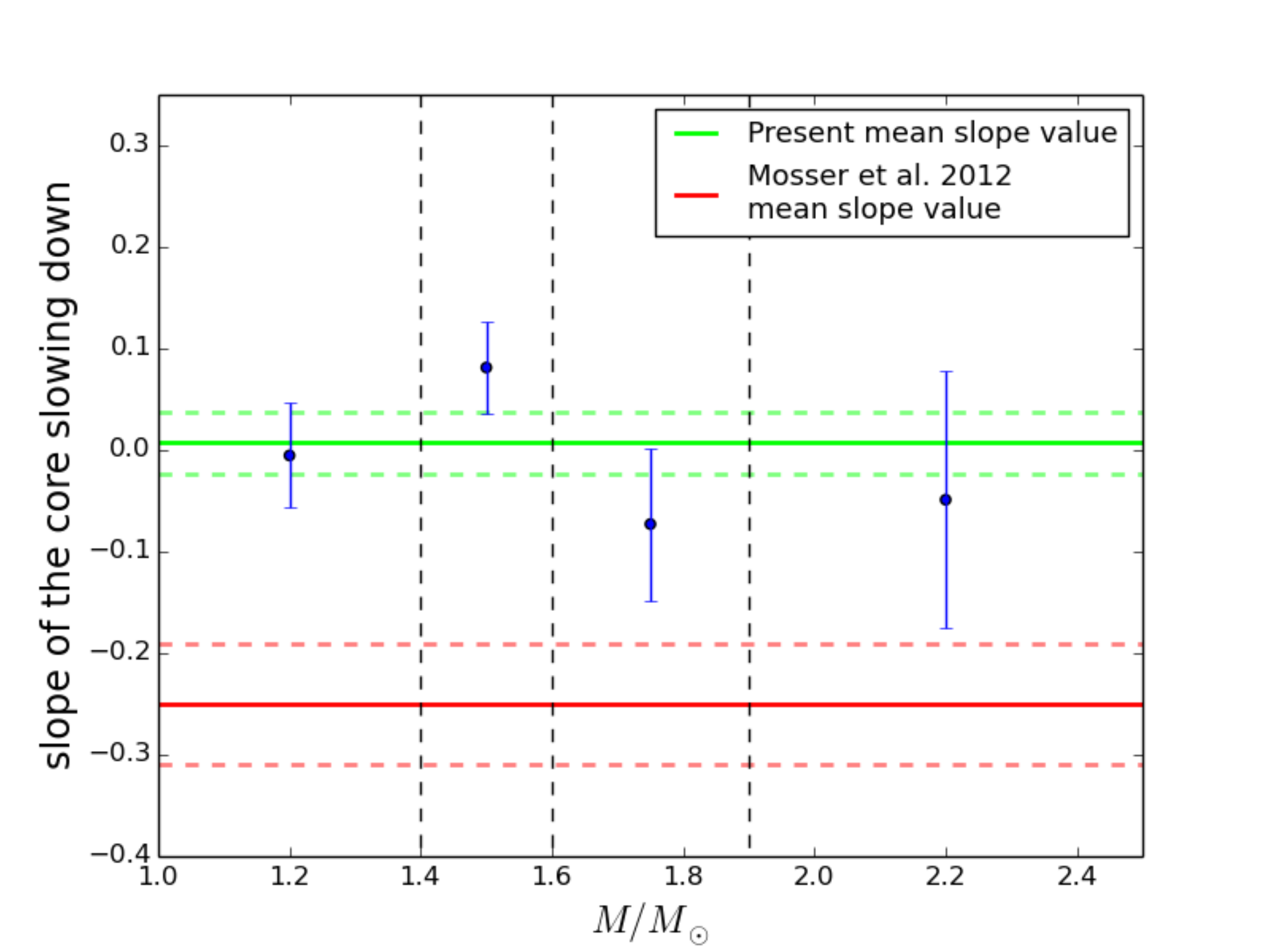}
\caption{Slopes of the core braking when considering the evolution of the core rotation as a function of the mixed-mode density for different mass ranges. Our measurements are represented in blue with the associated error bars. Vertical black dashed lines represent the boundaries between the different mass ranges considered. The green continuous and dashed lines indicate the mean value of the braking efficiency measured in this study and the associated error bars, respectively. The red continuous and dashed lines indicate the mean braking efficiency and the associated error bars estimated from \cite{Mosser_2012c} measurements, respectively.}
\label{fig-slopes}
\end{figure*}

We applied our method to 1714 RGB stars presenting a large mass range from 1 $\Msol$ up to 2.5 $\Msol$ \citep{Gehan_2017}. Stellar masses were estimated from the asteroseismic global parameters $\Dnu$ and $\numax$ \citep{Kjeldsen, Mosser_2013} through
\begin{equation}
\frac{M}{\Msol} = \left(\frac{\numax}{\numaxsol}\right)^3 \left(\frac{\Dnu}{\Dnusol}\right)^{-4} \left(\frac{\Teff}{\Tsol}\right)^{3/2},
\end{equation}
where $\Dnu$ is the large separation of pure pressure modes, and $\numax$ is the frequency of maximum oscillation power. $\numaxsol = 3050 \, \mu$Hz, $\Dnusol = 135.5 \, \mu$Hz and $\Tsol = 5777 \, K$ are the solar values chosen as references.
\newline
 Our method led to the identification of the rotational components for 69\% of cases. As it is not possible to retrieve information on the core rotation of stars seen pole-on, we obtained 875 mean core rotation measurements,
 with a typical uncertainty of about 10 nHz (Fig.~\ref{fig-dnurot}).
\newline
We considered the number of gravity modes per $\Dnu$-wide frequency range as a proxy of stellar evolution \citep{Gehan_2017}. This mixed-mode density is defined as
\begin{equation}\label{eqt-nmix}
    \N = {\Dnu \over \DPi \, \numax^2}.
\end{equation}
The large dataset considered in this study covering an extended mass range allowed us to investigate how the rate of the core braking depends on the stellar mass. We thus selected stars in four different mass ranges and fitted a relonship of the type
\begin{equation}\label{eqt-mass-range}
\dnurotcore \propto \N^{a}.
\end{equation}
The exponents $a$ are summarized in Table~\ref{table:1} as a function of the mass range considered.
The present result indicates that the braking rate of the core rotation does not seem to depend on the stellar mass (Fig.~\ref{fig-slopes}). When applying Eq.\ref{eqt-mass-range} to the measurements for RGB stars made by \cite{Mosser_2012c}, we obtain a mean braking rate much higher than the value measured in this study (Table~\ref{table:2}).
This discrepancy is mainly due to sample effects. Indeed, we now have a RGB sample about 10 times larger than \cite{Mosser_2012c}, thus more significant from a statistical point of view.
 The result obtained provides a more precise estimate of the slope of the core braking (Fig.~\ref{fig-slopes}). Nevertheless, the lower slope of the core braking measured in this study does not call into question the results of \cite{Mosser_2012c} predicting an efficient angular momentum transport inside red giant cores.

\begin{table}
\caption{Slope of the core braking as a function of the mixed-mode density $\N$ for different mass ranges}
\label{table:1}
\centering
\begin{tabular}{c c}
\hline\hline
$M$ & $a$ \\
\hline
$M \leq 1.4 \Msol$ & $- \, 0.01$ $\pm$ $0.05$ \\
$1.4 < M \leq 1.6 \Msol$ & $0.08$ $\pm$ $0.04$ \\
$1.6 < M \leq 1.9 \Msol$ & $- \, 0.07$ $\pm$ $0.07$ \\
$M > 1.9 \Msol$ & $- \, 0.05$ $\pm$ $0.13$ \\
\hline
\end{tabular}
\end{table}

\begin{table}
\caption{Slope of the core braking as a function of the mixed-mode density $\N$ for \cite{Mosser_2012c} red giant branch stars. $a\ind{2012}$ is the slope estimated from \cite{Mosser_2012c} measurements, $a$ is the slope measured in this study.}
\label{table:2}
\centering
\begin{tabular}{c c c}
\hline\hline
$a\ind{2012}$ & $a$\\
\hline
$- \, 0.25$ $\pm$ $0.06$ & $0.01$ $\pm$ $0.03$ \\
\hline
\end{tabular}
\end{table}

\section{Conclusions}
%%--------------------

Disentangling rotational splittings from mixed modes is now possible in a simple way, opening the era of large-scale measurements of the core rotation of red giants. We developed a method allowing an automated identification of the dipole gravity-dominated mixed-modes split by rotation, providing us with a measurement of the mean core rotational splitting $\dnurotcore$. The method satisfactorily identified the rotational components of 1181 red giant branch stars, representing a success rate of 69\% and providing us with 875 $\dnurotcore$ measurements. This large dataset including stars with masses as high as 2.5 $\Msol$ allowed us to show that the slope of the core braking does not seem to depend on the stellar mass, but also that this slope is much lower than what was estimated by Mosser et al. (2012). Nevertheless, this measurement does not go against a very efficient angular momentum transport inside red giant cores. This result suggests that the efficiency of the mechanism transporting angular momentum inside red giant branch stars should increase with the stellar mass in order to compensate simultaneously the stronger differential rotation which is expected to develop in higher mass stars and their faster temporal evolution.  Further investigation on the efficiency of the angular momentum transport inside red giants is necessary, through modelling of the evolution of the core inertia depending on the stellar mass.

\begin{acknowledgements}
We acknowledge financial support from "Programme National de Physique Stellaire" (PNPS) of CNRS/INSU, France.
\end{acknowledgements}

\bibliographystyle{aa}  % A&A bibliography style file (aa.bst)
\bibliography{biblio} % your references in file: Yourfile.bib

\begin{thebibliography}{16}
\expandafter\ifx\csname natexlab\endcsname\relax\def\natexlab#1{#1}\fi

\bibitem[{{Beck} {et~al.}(2011){Beck}, {Bedding}, {Mosser}, {Complement},
  {Complement}, \& {Complement}}]{Beck}
{Beck}, P.~G., {Bedding}, T.~R., {Mosser}, B., {et~al.} 2011, Science, 332, 205

\bibitem[{{Belkacem} {et~al.}(2015{\natexlab{a}}){Belkacem}, {Marques},
  {Goupil}, {Complement}, {Complement}, \& {Complement}}]{Belkacem_2015a}
{Belkacem}, K., {Marques}, J.~P., {Goupil}, M.~J., {et~al.} 2015{\natexlab{a}},
  \aap, 579, A30

\bibitem[{{Belkacem} {et~al.}(2015{\natexlab{b}}){Belkacem}, {Marques},
  {Goupil}, {Complement}, {Complement}, \& {Complement}}]{Belkacem_2015b}
{Belkacem}, K., {Marques}, J.~P., {Goupil}, M.~J., {et~al.} 2015{\natexlab{b}},
  \aap, 579, A31

\bibitem[{{Cantiello} {et~al.}(2014){Cantiello}, {Mankovich}, {Bildsten},
  {Complement}, {Complement}, \& {Complement}}]{Cantiello}
{Cantiello}, M., {Mankovich}, C., {Bildsten}, L., {et~al.} 2014, \apj, 788, 93

\bibitem[{{De Ridder} {et~al.}(2009){De Ridder}, {Barban}, {Baudin},
  {Complement}, {Complement}, \& {Complement}}]{De_Ridder}
{De Ridder}, J., {Barban}, C., {Baudin}, F., {et~al.} 2009, Nature, 459, 398

\bibitem[{{Fuller} {et~al.}(2014){Fuller}, {Lecoanet}, {Cantiello},
  {Complement}, {Complement}, \& {Complement}}]{Fuller}
{Fuller}, J., {Lecoanet}, D., {Cantiello}, M., {et~al.} 2014, \apj, 796, 17

\bibitem[{{Gehan} {et~al.}(2016{\natexlab{a}}){Gehan}, {Mosser}, \&
  {Michel}}]{Gehan_2016a}
{Gehan}, C., {Mosser}, B., \& {Michel}, E. 2016{\natexlab{a}}, ArXiv e-prints

\bibitem[{{Gehan} {et~al.}(2016{\natexlab{b}}){Gehan}, {Mosser}, \&
  {Michel}}]{Gehan_2016b}
{Gehan}, C., {Mosser}, B., \& {Michel}, E. 2016{\natexlab{b}}, ArXiv e-prints

\bibitem[{{Gehan} {et~al.}(2018){Gehan}, {Mosser}, \& {Michel}}]{Gehan_2017}
{Gehan}, C., {Mosser}, B., \& {Michel}, E. 2018, in prep.

\bibitem[{{Goupil} {et~al.}(2013){Goupil}, {Mosser}, {Marques}, {Complement},
  {Complement}, \& {Complement}}]{Goupil}
{Goupil}, M.~J., {Mosser}, B., {Marques}, J.~P., {et~al.} 2013, A\&A, 549, A75

\bibitem[{{Kjeldsen} \& {Bedding}(1995)}]{Kjeldsen}
{Kjeldsen}, H. \& {Bedding}, T.~R. 1995, \aap, 293, 87

\bibitem[{{Lagarde} {et~al.}(2012){Lagarde}, {Decressin}, {Charbonnel},
  {Eggenberger}, {Ekstr{\"o}m}, \& {Palacios}}]{Lagarde}
{Lagarde}, N., {Decressin}, T., {Charbonnel}, C., {et~al.} 2012, \aap, 543,
  A108

\bibitem[{{Mosser} {et~al.}(2012){Mosser}, {Goupil}, {Belkacem}, {Complement},
  {Complement}, \& {Complement}}]{Mosser_2012c}
{Mosser}, B., {Goupil}, M.~J., {Belkacem}, K., {et~al.} 2012, A\&A, 548, A10

\bibitem[{{Mosser} {et~al.}(2013){Mosser}, {Michel}, {Belkacem}, {Complement},
  {Complement}, \& {Complement}}]{Mosser_2013}
{Mosser}, B., {Michel}, E., {Belkacem}, K., {et~al.} 2013, \aap, 550, A126

\bibitem[{{Mosser} {et~al.}(2015){Mosser}, {Vrard}, {Belkacem}, {Complement},
  {Complement}, \& {Complement}}]{Mosser_2015}
{Mosser}, B., {Vrard}, M., {Belkacem}, K., {et~al.} 2015, A\&A, 584, A50

\bibitem[{{Pin{\c c}on} {et~al.}(2017){Pin{\c c}on}, {Belkacem}, {Goupil},
  {Complement}, {Complement}, \& {Complement}}]{Pincon}
{Pin{\c c}on}, C., {Belkacem}, K., {Goupil}, M.~J., {et~al.} 2017, ArXiv
  e-prints

\end{thebibliography}

\end{document}